\documentstyle[12pt]{article}

\font\tenrm=cmr10

 1
 1
 1

\def\ie{{\em i.e.}}
\def\eg{{\em e.g.}}

\def\beq{\begin{equation}}
\def\eeq{\end{equation}}
\def\bdm{\begin{displaymath}}
\def\edm{\end{displaymath}}

\catcode`\@=11 
\def\coeff#1#2{{\textstyle{#1\over #2}}}

\def\vev#1{\left\langle #1\right\rangle}
\def\lsim{\mathrel{\mathpalette\@versim<}}
\def\gsim{\mathrel{\mathpalette\@versim>}}
\def\@versim#1#2{\vcenter{\offinterlineskip
    \ialign{$\m@th#1\hfil##\hfil$\crcr#2\crcr\sim\crcr } }}
\def\etal{{\em et. al.}}
\def\JL{J. L. Lopez}
\def\DVN{D. V. Nanopoulos}

\def\r#1{$\bf#1$}
\def\rb#1{$\bf\overline{#1}$}

\def\t1{{\tilde 1}}

\def\GeV{\,{\rm GeV}}
\def\TeV{\,{\rm TeV}}
\def\y{\,{\rm y}}

\def\wt{\widetilde}

\def\to{\rightarrow}

\def\NPB#1#2#3{Nucl. Phys. B {\bf#1} (19#2) #3}
\def\PLB#1#2#3{Phys. Lett. B {\bf#1} (19#2) #3}

\def\PRD#1#2#3{Phys. Rev. D {\bf#1} (19#2) #3}
\def\PRL#1#2#3{Phys. Rev. Lett. {\bf#1} (19#2) #3}
\def\PRT#1#2#3{Phys. Rep. {\bf#1} (19#2) #3}
\def\MODA#1#2#3{Mod. Phys. Lett. A {\bf#1} (19#2) #3}

\def\TAMU#1{Texas A \& M University preprint CTP-TAMU-#1}

\begin{document}
\begin{flushright}
\baselineskip=12pt
{CTP-TAMU-42/93}\\
{ACT-16/93}\\
\end{flushright}

\begin{center}
\vglue 2cm
{\Large\bf SUSY GUTs: A Practical Introduction}
\footnote{To appear in the Proceedings of the International School of
Subnuclear Physics 31st Course: ``From Supersymmetry to the Origin of
Space-Time", Erice, Italy, July 4-12, 1993.}\\
\vglue 1cm
{JORGE L. LOPEZ\\}
\vglue 0.5cm
{\em Center for Theoretical Physics, Department of Physics, Texas A\&M
University\\}
{\em College Station, TX 77843--4242, USA\\}
\baselineskip=12pt
\vglue 1cm
{\tenrm ABSTRACT}
 \end{center}
{\rightskip=3pc
 \leftskip=3pc
\noindent
An introduction to the most important concepts in the subject of supersymmetric
unified theories is presented. The emphasis is on the practical aspects leading
to state-of-the-art calculations in this renascent subject. The topics covered
include: generalities of supersymmetric unified theories, gauge and Yukawa
coupling unification including the most up-to-date numerical analyses, soft
supersymmetry breaking, and radiative electroweak symmetry breaking enforced
using the tree-level and one-loop effective potentials. This class of
supersymmetric models can be described in terms of five parameters: the
top-quark mass ($m_t$), the ratio of Higgs vacuum expectation values
($\tan\beta$), and three universal soft-supersymmetry-breaking parameters
($m_{1/2},m_0,A$). Thus, highly correlated predictions can be expected for all
conceivable experimental observables. In effect, these general models provide a
basic framework upon which more constrained models can be built.}
\vspace{2cm}
\begin{flushleft}
\baselineskip=12pt
{CTP-TAMU-42/93}\\
{ACT-16/93}\\
July 1993
\end{flushleft}
\vfill\eject
\tableofcontents
\vfill\eject
\setcounter{page}{1}
\pagestyle{plain}

\baselineskip=14pt

\section{What are SUSY GUTs?}
I will define a supersymmetric unified theory as one that incorporates the
following elements in one guise or another:
\begin{description}
\item (i) {\em GUT Symmetry}: which manifests itself in the unification of the
{\em non-abelian} gauge couplings of the standard model above certain
``unification scale" ($M_U$). Gauge groups with this property include: $SU(5)$,
$SO(10)$, and $SU(5)\times U(1)$.
\item (ii) {\em GUT Fields}: which only exist because the larger GUT symmetry
is present and which decouple below the unification scale. These fields play
important roles in the traditional GUT processes such as gauge symmetry
breaking, doublet-triplet (2/3) splitting, proton decay, baryogenesis, neutrino
masses, etc.
\item (iii) {\em Yukawa Unification}: the larger gauge symmetries force the SM
fermions to ``share" larger representations, and therefore the gauge invariant
Yukawa couplings in the GUT phase usually encompass more than one ``low-energy"
Yukawa coupling. Specifically one usually gets the following relations valid at
$M_U$:
\beq
\begin{array}{ll}
\lambda_b=\lambda_\tau&SU(5),\\
\lambda_b=\lambda_\tau=\lambda_t&SO(10),\\
\lambda_t=\lambda_{\nu_\tau}&SU(5)\times U(1).
\end{array}
\eeq
\item (iv) {\em (Universal) Soft-Supersymmetry Breaking}: spontaneous breaking
of supergravity yields a global supersymmetric theory supplemented by a set
of soft-supersymmetry-breaking parameters. If these parameters are $\lsim{\cal
O}(1\TeV)$, supersymmetry effectively solves the gauge hierarchy problem.
\item (v) {\em Dynamical Evolution}: renormalization group equations (RGEs) for
the gauge, Yukawa, and scalar couplings relate the values of these parameters
at the high- and low-energy scales.
\item (v$1\over2$) {\em Intermediate Scale Fields}: are apparently needed for
string unification.
\item (vi) {\em Light Fields}: should include the standard model fields with
two Higgs doublets plus their superpartners, and maybe other light fields
such as Higgs singlets.
\item (vii) {\em Radiative Electroweak Symmetry Breaking}: allows the
generation of the electroweak scale dynamically. The top-quark mass plays an
important role.
\end{description}

The theories outlined above have the most appealing property of requiring
at most {\em five} parameters to describe all new phenomena (excluding new
sources of CP violation). These parameters are:
\begin{description}
\item[--] The top-quark mass ($m_t\gsim110\GeV$), to be measured soon at the
Tevatron;
\item[--] The ratio of Higgs vacuum expectaction values (VEVs)
($ 1<\tan\beta\lsim50$);
\item[--] Three soft-supersymmetry breaking parameters: the universal gaugino
mass ($m_{1/2}\propto m_{\tilde g}$), the universal scalar mass ($m_0$), and
the universal cubic scalar coupling ($A$).
\end{description}
This relative scarcity of parameters is in sharp contrast with the more than
twenty parameters needed to achieve a compararable description in the minimal
supersymmetric standard model (MSSM). An immediate consequence is that {\em
all} sparticle and Higgs boson masses and couplings can be determined in terms
of these five parameters, and therefore lots of non-trivial and unsuspected
correlations arise. As such, these theories provide a very predictive scenario
for \eg, collider processes and rare decays.

An important remark to keep in mind is that it is essential to consider {\bf
all} aspects of the models under consideration, \eg:
\begin{description}
\item[--] SU(5): Gauge and Yukawa unification have been emphasized vigorously
in the recent literature
\cite{EKNI,LL,Arason,AnselmoI,EKNII,EKNIII,EKNIV,RR,HMY,
AnselmoIII,AnselmoIV,AnselmoVI,LP,CPW,BBO,BH,HS}. {\em However}, proton decay
is a very important constraint  \cite{ENR,ANoldpd,MATS,ANpd,HMY,LNP,LNPZ} and
so is cosmology \cite{troubles,LNP,ANcosm,poles}, but these aspects of the
model have not received the same degree of attention (perhaps because of the
perception that they entail ``model-dependent" assumptions, although in
practice no new unknowns are introduced). The doublet-triplet (2/3) splitting
problem also needs to be addressed in a consistent way. For example, how does
the  missing-partner-mechanism (MPM) \cite{MPM} solution affect the GUT
threshold corrections \cite{HY}?
\item[--] SO(10): Predicts large $m_t$ and $\tan\beta$. Can one have radiative
electroweak breaking? what about proton decay? or the 2/3 splitting problem?
\end{description}

\section{Gauge Coupling Unification}
\subsection{Generalities}
\begin{description}
\item[--]Traditional GUTs: (grand) unified theories (\eg, $SU(5),SO(10),E_6$,
and $SU(5)\times U(1)$) generally contain:
\begin{description}
\item[--] Larger/new structure, revealed above some ``unification" scale;
\item[--] Some sort of gauge and Yukawa coupling unification;
\item[--] Observable proton decay, baryogenesis, neutrino masses, etc.
\end{description}
Examples of {\em non} GUTs include $SU(3)\times SU(2)_L\times U(1)_Y$ (SM) and
$SU(4)\times SU(2)_L\times SU(2)_R$ (Pati-Salam)
\item[--] GUSTs: in grand unified {\em superstring} theories all the above
generic properties are realized in sometimes novel ways:
\begin{description}
\item[--] Larger structure is provided by string massive states;
\item[--] Gauge unification is automatic in this top$\to$down approach (in
simple models), and it is understandable in terms of a ``primordial $SO(44)$"
gauge symmetry;
\item[--] Yukawa unification happens in a disorderly way as a consequence of
remnants of higher symmetries \cite{decisive}.
\end{description}
\item[--] Intermediate Scales?

Symmetry breaking patterns can include intermediate scales or not:\\

\noindent $SU(3)\times SU(2)\times U(1)\to SU(5)$ (minimal GUT unification)\\

\noindent $SU(3)\times SU(2)\times U(1)\to SO(10)$ (one-step unification)\\
$SU(3)\times SU(2)\times U(1)\to SU(5)\to SO(10)$ (two-step unification)\\

\noindent $SU(3)\times SU(2)\times U(1)\to SU(5)\times U(1)\to  String$
(intermediate unification)\\
$SU(3)\times SU(2)\times U(1)\to String \,(SU(5)\times U(1))$ (one-step string
unification)\footnote{It appears that $SU(5)\times U(1)$ only makes sense as a
unified theory in the context of strings, since otherwise beyond the string
scale the $SU(5)$ and $U(1)$ couplings would diverge again.}\\
\item[--] Non-minimal Matter: for example, $Q,\bar Q$, $D^c,\bar D^c$ allow
$SU(3)\times SU(2)\times U(1)$ to unify at $\sim10^{18}\GeV$ if their masses
are chosen appropriately \cite{price}. These fields fit snugly in the
$SU(5)\times U(1)$ representations \cite{sism}:
\beq
{\bf10}=\{Q,D^c,\nu^c\},\qquad \overline{\bf10}=\{\bar Q,\bar D^c,\bar\nu^c\}.
\eeq
\end{description}

\subsection{Renormalization Group Equations (RGEs)}

If unification is assumed to occur in a single step and there are no
intermediate scale particles with poorly determined masses, then one can
study this problem in great detail (\eg, in $SU(5)$)\\

\noindent{\em Problem:} solve the coupled set of two-loop gauge and Yukawa
coupling RGEs, taking into account low- and high-energy threshold effects.\\

\noindent{\em Objective:} To obtain $\alpha_U$, $M_U$, and $\sin^2\theta_W$ in
terms of $\alpha_e$ and $\alpha_3$ (both measured at $M_Z$), the light
supersymmetric spectrum,  and the heavy GUT spectrum.\\

\noindent{\em Philosophical Note}:
\begin{description}
\item[--] Our approach assumes that gauge coupling unification occurs, as is
the case in a unified theory. The model is tested by comparing its prediction
for $\sin^2\theta_W$ against the experimental value. This is the top$\to$down
approach.
\item[--] Compare this with the (``experimental") bottom$\to$up approach where
one ``tests" for unification by running up the gauge couplings. What can one
conclude if it is found that the couplings do not meet?
\end{description}

\bigskip
\bigskip
\noindent The gauge coupling RGEs are given by
\begin{equation}
{{dg_i}\over {dt}}={g_i\over{16\pi^2 }}\left [b_ig_i^2+{1\over {16\pi^2 }}
\left (\sum _{j=1}^3b_{ij}g_i^2g_j^2-\sum _{j=t,b,\tau}a_{ij}g_i^2
\lambda _j^2\right )\right ], \label{dgidt}
\end{equation}
where $t=\ln (Q/M_U)$ with $Q$ the running scale and $M_U$ the unification
mass. Also, $\alpha_1={5\over3}(\alpha_e/\cos^2\theta_W)$,
$\alpha_2=(\alpha_e/\sin^2\theta_W)$, and

\begin{eqnarray}
b_i&=&\left(\coeff{33}{5},1,-3\right), \\
b_{ij}&=&\left( \begin{array}{c@{\quad}c@{\quad}c}
{199\over 25} & {27\over 5} & {88\over 5} \\
{9\over 5} & 25 & 24 \\ {11\over 5} & 9 & 14
\end{array} \right),\\
a_{ij}&=&\left( \begin{array}{c@{\quad}c@{\quad}c}
{26\over 5} & {14\over 5} & {18\over 5} \\
6 & 6 & 2 \\ 4 & 4 & 0
\end{array} \right).
\end{eqnarray}

\subsection{Analytic solutions}

Neglecting the very small effect of the Yukawa couplings in the gauge coupling
evolution \cite{BBO}, one can write down analytic solutions to the RGEs to the
desired accuracy (see \eg, \cite{AnselmoIII}). First we neglect all heavy
thresholds, \ie, we assume unification occurs at one point. The solutions are:
\begin{eqnarray}
\ln{M_U\over M_Z}&=&{\pi\over10}\left({1\over\alpha_e}-{8\over3\alpha_3}\right)
		\quad\qquad\qquad{\rm (one-loop)}	\nonumber\\
&&-{1\over40}\left(C_2+\coeff{5}{3}C_1-\coeff{8}{3}C_3\right)
	\qquad\!{\rm (two-loop)}\nonumber\\
&&+\sum_i p_i\ln{\wt m_i\over M_Z}
\qquad\qquad\qquad{\rm (light\ thresholds)}\label{MU}\\
%
{\alpha_e\over\alpha_U}&=&{3\over20}\left(1+{4\alpha_e\over\alpha_3}\right)
		\nonumber\\
&&+{\alpha_e\over80\pi}\left[b_3 C_2+\coeff{5}{3}b_3C_1-\left(b_2+\coeff{5}{3}
b_1\right)C_3\right]
	\nonumber\\
&&+{\alpha_e\over2\pi}\sum_i q_i\ln{\wt m_i\over M_Z}\label{alphaU}\\
%
\sin^2\theta_W&=&0.2+{7\alpha_e\over15\alpha_3}\nonumber\\
&&-{5\over3}{\alpha_e\over80\pi}\left[(b_1-b_2)C_3+(b_3-b_1)C_2+(b_2-b_3)C_1
\right]\nonumber\\
&&+{\alpha_e\over20\pi}\sum_i r_i\ln{\wt m_i\over M_Z}\nonumber\\
&&+ \Delta T_H	\qquad\qquad\qquad\qquad{\rm (heavy\ thresholds)}\label{sin2}
\end{eqnarray}
The $p_i,q_i,r_i$ coefficients are given in Table~\ref{Table1}. Also,
\beq
C_i=\sum_j{b_{ij}\over b_j}\ln{\alpha^{-1}_j(M_Z)\over\alpha^{-1}_U}
+\left(\sum_j{b'_{ij}\over b'_j}-\sum_j{b_{ij}\over b_j}\right)
\ln{\alpha^{-1}_j(M_Z)\over\alpha^{-1}_j(\widetilde m)}
\eeq
with the $b'_i,b'_{ij}$ the one- and two-loop non-supersymmetric RGE
coefficients,
\begin{eqnarray}
b'_i&=&\left(\coeff{41}{10},-\coeff{19}{6},-7\right), \\
b'_{ij}&=&\left( \begin{array}{c@{\quad}c@{\quad}c}
{199\over 50} & {27\over 10} & {44\over 5} \\
{9\over 10} & {35\over6} & 12 \\ {11\over 10} & {9\over2} & -26
\end{array} \right).
\end{eqnarray}

\begin{table}[t]
\hrule
\caption{The coefficients which weigh the light threshold corrections to the
unification mass ($p_i$), the unified coupling ($q_i$), and $\sin^2\theta_W$
($r_i$).}\label{Table1}
\begin{center}
\begin{tabular}{|c|c|c|c|}\hline
$i$&$p_i$&$q_i$&$r_i$\\ \hline
$t$&$\coeff{1}{120}$&$\coeff{83}{120}$&$-3$\\
$\widetilde w$&$\coeff{1}{15}$&$\coeff{1}{5}$&$-\coeff{32}{3}$\\
$\tilde g$&$-\coeff{4}{15}$&$\coeff{6}{5}$&$-\coeff{28}{3}$\\
$\tilde h$&$\coeff{1}{15}$&$\coeff{1}{5}$&$-4$\\
$H$&$\coeff{1}{60}$&$\coeff{1}{20}$&$-1$\\
$\tilde q$&$-\coeff{5}{48}$&$\coeff{65}{48}$&$\coeff{5}{2}$\\
$\tilde t_L$&$\coeff{1}{240}$&$\coeff{43}{240}$&$-\coeff{19}{6}$\\
$\tilde t_R$&$0$&$\coeff{1}{6}$&$\coeff{5}{3}$\\
$\tilde l_L$&$\coeff{1}{20}$&$\coeff{3}{20}$&$-3$\\
$\tilde l_R$&$\coeff{1}{20}$&$\coeff{3}{20}$&$2$\\ \hline
\end{tabular}
\end{center}
\hrule
\end{table}

\noindent{\em Note:} In these equations the non-supersymmetric regime includes
only the lighter Higgs doublet. The symbol $\widetilde m$ in the definition of
the $C_i$ coefficients is an average sparticle mass.\\

\noindent$\bullet$ {\bf Two comments about threshold effects:}
\begin{description}
\item[--] The sparticles are decoupled in a single-step approximation at a mass
scale equal to their physical mass in both the $\overline{MS}$ and
$\overline{DR}$ schemes used to treat the light and heavy sectors of the theory
respectively \cite{EKNIII,LP}. (Exception: in the $\overline{MS}$ scheme the
spin-1 particles are decoupled at $e^{-1/21}\approx0.95$ of their mass.)
\item[--] Note in the above equations (\ref{MU},\ref{alphaU},\ref{sin2})
that the threshold effects are comparable in size to the two-loop effects.
\end{description}

\noindent$\bullet$ {\bf Heavy Thresholds:}
If symmetry breaking occurs because of the VEV of the \r{24} of Higgs, then
only three GUT masses are needed to parametrize the relevant effects:
(i) the masses of the $X,Y$ gauge bosons $M_V$, (ii) the mass of the adjoint
Higgs multiplet $M_\Sigma$, and (iii) the mass of the color triplet Higgs
fields $M_H$ (which mediate proton decay). The contribution to $\sin^2\theta_W$
is given by \cite{EKNIII,BH,HMY}
\beq
\Delta T_H={\alpha_e\over20\pi}\left(-6\ln{M_U\over M_H}+4\ln{M_U\over M_V}
+2\ln{M_U\over M_\Sigma}\right),
\eeq
where $M_U$ is the largest of the three masses, \ie, a ``unification
scale" does not exist. Since proton decay requires a large $M_H$, most likely
$\Delta T_H>0$.\\

\subsection{Numerical Status}
Initially it was thought possible to determine the supersymmetry scale by a
``best fit" to unification. Early studies even claimed that the supersymmetric
spectrum had to lie in the TeV region to possibly achieve unification. However,
it was eventually realized that several uncertainties in the calculations (most
notably the heavy GUT thresholds) \cite{BH,AnselmoIV,EKNIV} do not allow to
constrain the supersymmetric parameters more than within a few TeV, that is,
there is no real constraint on the supersymmetric particle masses from these
analyses. Implementing the objective described above, one finds that by varying
all parameters in the calculation one  obtains $\sin^2\theta_W$ within the
experimental range (see \eg, \cite{EKNIII,LP}), although certain combinations
of the parameters are not allowed. The most up-to-date analysis is by Langacker
and Polonsky~\cite{LP}:
\begin{description}
\item[--] Experimental data:
\begin{eqnarray}
M_Z&=&91.187\pm0.007\GeV\\
\alpha^{-1}_e&=&127.9\pm0.1\\
\alpha_3(M_Z)&=&0.120\pm0.010
\end{eqnarray}
\item[--] Two-parameter fit to all $W^\pm$,$Z$, and neutral current data:
\begin{eqnarray}
\sin^2\theta_W&=&0.2324\pm0.0006\label{sin2exp}\\
m_t&=&138^{+20}_{-25}+5\GeV\label{mtexp}
\end{eqnarray}
where $+5$ is due to the supersymmetric Higgs variation ($m_h=50-150\GeV$)
\item[--] Varying {\em all} parameters, $SU(5)$ GUT gives\footnote{Note:
$0.2334=\underbrace{0.2304}_{\rm one-loop}+\underbrace{0.0030}_{\rm two-loop}$,
thus threshold effects are comparable to two-loop effects.}
\begin{eqnarray}
\sin^2\theta_W&=&0.2334\quad\qquad(m_t=138, m_h=\widetilde m=M_Z)\nonumber\\
&&\pm0.0025\qquad(\alpha_e,\alpha_3)\nonumber\\
&&\pm0.0014\qquad({\rm light\ thresholds})\nonumber\\
&&\pm0.0006\qquad(m_t,m_h)\nonumber\\
&&{}^{+0.0013}_{-0.0005}\quad\qquad({\rm heavy\ thresholds})\label{sin2th}
\end{eqnarray}
The individual contributions to each of these effects are shown in
Fig.~\ref{Figure1} \cite{LP}. Since all of the parameters inducing
uncertainties in the predicted value of $\sin^2\theta_W$ are independent,
contrasting the prediction in Eq.~(\ref{sin2th}) with the experimental
determination in Eq.~(\ref{sin2exp}), one can rule out some combinations of the
various parameters. However, the various uncertainties appear to be too large
to make any definite statements. Alternatively, to cirmcumvent the large
uncertainty on $\alpha_3$, one can input $\sin^2\theta_W$ and obtain the
predicted value of $\alpha_3$
\beq
\alpha_3(M_Z)=0.125\pm0.001\pm0.005\pm0.002^{+0.005}_{-0.002}
\eeq
where the first error is now due to $\sin^2\theta_W$ and the others are as
above. Here again, it is clear that gauge coupling unification in the minimal
$SU(5)$ supergravity model is in very good agreement with low-energy data.
\end{description}

\begin{figure}[p]
\caption{Contributions from individual correction terms to the $SU(5)$
GUT prediction for $\sin^2\theta_W$ (from Ref. [13]). Dashed line: error bar on
$\sin^2\theta_W$. Dashed-dotted line: uncertainty induced by $\alpha_3$ on the
prediction for $\sin^2\theta_W$. Dotted line: two-loop contribution to
$\sin^2\theta_W$.}\label{Figure1}
\vspace{7.5in}
\end{figure}

\section{Yukawa Coupling Unification}
We now consider the further constrain where two or three of the
third-generation Yukawa couplings are unified at the scale $M_U$. For
completeness, first we present the relevant two-loop RGEs (from \cite{BBO})
which should be used in conjunction with the gauge coupling RGEs in the
previous section.
\subsection{Two-loop RGEs}
\begin{eqnarray}
{{d\lambda _t}\over {dt}}={{\lambda _t}\over {16\pi ^2}}
\Bigg [\Bigg (&-&\sum_i c_ig_i^2+6\lambda _t^2+\lambda _b^2
\Bigg )
\nonumber \\
&+&{1\over {16\pi ^2}}\Bigg (\sum_i \left(c_ib_i+c_i^2/2\right )g_i^4
+g_1^2g_2^2+{136\over 45}g_1^2g_3^2+8g_2^2g_3^2\nonumber \\
&&\;\;\;\;\;\;
+\lambda _t^2\left({6\over 5}g_1^2+6g_2^2+16g_3^2\right )+{2\over
5}\lambda_b^2g_1^2 \nonumber \\
&&\;\;\;\;\;\;
-\left \{22\lambda _t^4+5\lambda _t^2\lambda _b^2
+5\lambda _b^4+\lambda _b^2\lambda _{\tau}^2\right \}\Bigg )\Bigg ]
\label{dytdt}
\end{eqnarray}
\begin{eqnarray}
{{d\lambda _b}\over {dt}}={{\lambda _b}\over {16\pi ^2}}
\Bigg [\Bigg (&-&\sum_i c_i^{\prime}g_i^2+\lambda _t^2
+6\lambda _b^2+\lambda _{\tau}^2
\Bigg )
\nonumber \\
&+&
{1\over {16\pi ^2}}\Bigg (\sum_i \left(c_i^{\prime}b_i+c_i^{\prime 2}
/2\right )g_i^4+g_1^2g_2^2+{8\over 9}g_1^2g_3^2+8g_2^2g_3^2\nonumber \\
&&\;\;
+{4\over 5}\lambda_t^2g_1^2 +\lambda _b^2
\left({2\over 5}g_1^2+6g_2^2+16g_3^2\right )+{6\over 5}
\lambda_{\tau}^2g_1^2 \nonumber \\
&&\;\;
-\left \{22\lambda _b^4+5\lambda _t^2\lambda _b^2
+3\lambda _b^2\lambda _{\tau}^2+3\lambda _{\tau}^4+5\lambda _t^4
\right \}\Bigg )\Bigg ]
\label{dybdt}
\end{eqnarray}
\begin{eqnarray}
{{d\lambda _{\tau}}\over {dt}}={{\lambda _{\tau}}\over {16\pi ^2}}
\Bigg [\Bigg (&-&\sum_i c_i^{\prime \prime}g_i^2
+3\lambda _b^2+4\lambda _{\tau}^2
\Bigg )
\nonumber \\
&+&
{1\over {16\pi ^2}}\Bigg (\sum_i \left(c_i^{\prime \prime}b_i
+c_i^{\prime \prime 2}/2\right )g_i^4
+{9\over 5}g_1^2g_2^2\nonumber \\
&&\;\;\;\;\;\;
+\lambda _b^2\left(-{2\over 5}g_1^2+16g_3^2\right )
+\lambda_{\tau}^2\left({6\over 5}g_1^2+6g_2^2\right ) \nonumber \\
&&\;\;\;\;\;\;
-\left \{3\lambda _t^2\lambda _b^2
+9\lambda _b^4+9\lambda _b^2\lambda _{\tau}^2+10\lambda _{\tau}^4
\right \}\Bigg )\Bigg ]
\label{dytaudt}
\end{eqnarray}
With
\beq
c_i=\left({13\over 15},3,{16\over 3}\right), \quad
c_i^{\prime}=\left({7\over 15},3,{16\over 3}\right),\quad
c_i^{\prime \prime}=\left({9\over 5},3,0\right).
\eeq

\subsection{Fixed points}
\label{fixed}
Independently of the GUT relations among the Yukawa couplings,
$\lambda_{b,t,\tau}$ must be bounded above at low energies (\ie,
$\lambda_{b,t,\tau}\lsim1$), otherwise they would blow up before reaching
$M_U$ (\ie, a Landau pole is encountered, see \eg, \cite{DL}). Using this
fact one can obtain an upper bound on the top-quark mass,
\begin{eqnarray}
m_t&=&\lambda_t v_2=\lambda_t{v_0\over\sqrt{2}}\sin\beta\nonumber\\
&&\lsim(174\GeV)\lambda^{max}_t{1\over\sqrt{1+1/\tan^2\beta}}\nonumber\\
&&\approx135,170,180,190\GeV\\
{\rm for\ }\tan\beta&=&1,2,3,\infty
\end{eqnarray}
The numerical upper bound ($\lambda^{max}_t\approx1.09$) depends on $\alpha_3$,
the light thresholds, etc \cite{BBO}. These $\tan\beta$-dependent upper bounds
are quite relevant nowadays, and could rule out a whole class of supersymmetric
unified theories, or more likely, provide (somewhat mild) lower bounds on
$\tan\beta$. Analogously, the bottom-quark Yukawa coupling upper bound entails
an upper bound on $\tan\beta$,
\begin{eqnarray}
m_b(M_Z)&=&{1\over\eta_b}m_b(m_b)=\lambda_b{v_0\over\sqrt{2}}
\cos\beta\nonumber\\
&&\lsim(174\GeV)\lambda^{max}_b{1\over\sqrt{1+\tan^2\beta}}\nonumber\\
\Rightarrow \tan\beta&\lsim&{190\eta_b\over m_b(m_b)}\approx50,55,58\\
{\rm for\ } m_b(m_b)&=&5.0,4.5,4.25\GeV
\end{eqnarray}
with $\eta_b=m_b(m_b)/m_b(M_Z)\approx1.3$.

\subsection{Unification conditions}
\begin{description}
\item (A) {\bf SU(5)}: the relation $\lambda_b(M_U)=\lambda_\tau(M_U)$, entails
a constraint on the $(m_t,\tan\beta)$ plane for given $m_b,\alpha_3$
\cite{EKNIII,YU,DHR,BBO,BBOII,LPII}. The procedure to determine this constraint
is somewhat complicated: (a) for a given $\tan\beta$ and $m_b(m_b)$ (and
$m_\tau$) one determines the low-energy values of $\lambda_{b,\tau}$;
(b) one runs these Yukawa coupling up to the given value of $m_t$, and
determines $\lambda_t(m_t)$; (c) then all three $\lambda_{b,t,\tau}$ are
run up to the unification scale and the GUT relation is tested; (d) the
given value of $m_t$ is adjusted until the GUT relation is satisfied. The
result of the calculations is a set of curves in the $(m_t,\tan\beta)$ plane
for fixed values of $m_b(m_b)$ and $\alpha_3$. A sample set of these curves
is shown in Fig.~\ref{Figure2} (taken from Ref.~\cite{BBO}) and show:
\begin{description}
\item[--] If $m_b=4.25\pm0.15\GeV$ (the shaded areas), then either: (i)
$\tan\beta\sim1$ or $\gsim40$, for a wide range of $m_t$ values, or (ii)
$m_t\gsim180\GeV$ for a wide range of $\tan\beta$ values. The value of
$\alpha_3$ has a moderate effect in this case. If $\alpha_3\gsim0.12$ then the
second possibility (\ie, $\tan\beta\gsim40$) is eliminated. It is important to
note that these predictions have been obtained without enforcing the gauge
coupling unification constraint to a high degree of precision, that is, the
points in the shaded areas do not necessarily give a value of $\sin^2\theta_W$
which is consistent with the experimentally allowed range. Enforcing this
constraint more precisely leads to much narrower allowed bands \cite{LPII}.
\item[--] If $m_b\approx5\GeV$ then the constraint on $\tan\beta$ is rather
weak and the value of $\alpha_3$ is quite relevant to the results
\cite{EKNIII,YU}.
\item[--] With what confidence can we say that $m_b\approx5\GeV$ is excluded?
The early discussions on this matter did not settle this issue satisfactorily;
subsequent discussions were effectively quelled by the Particle Data Group
published value of $m_b=4.25\pm0.15\GeV$. For a recent reappraisal see
Ref.~\cite{HRS}.
\item[--] What about GUT threshold effects that may correct this relation
somewhat? Allowing $\lambda_b<\lambda_\tau$ (see Fig.~\ref{Figure3}, from
Ref.~\cite{BBOII}) is equivalent to increasing $m_b$ and viceversa. For
example, solutions with $m_b\approx4.25\GeV$ would lead to much relaxed
constraints in the $(m_t,\tan\beta)$ plane if
$\lambda_b\approx0.8\lambda_\tau$.
\end{description}
\item (B) {\bf SO(10)}: the relation
$\lambda_b(M_U)=\lambda_\tau(M_U)=\lambda_t(M_U)$ is obtained in the simplest
$SO(10)$ GUT models and determines $m_t$ and $\tan\beta$ for given
$m_b,\alpha_3$ (\ie, one point on the $SU(5)$ curves in Fig.~\ref{Figure2}).
The procedure is similar to that described above for $SU(5)$ but the additional
relation allows $\tan\beta$ to be adjusted also. Because of this delicate
tuning, the results are quite sensitive to the various correction factors, but
generally give $m_t\gsim160\GeV$ and $\tan\beta\gsim40$ (see \eg,
\cite{ALS,Arason,YU,BBO,HRS}).
\end{description}

\begin{figure}[p]
\caption{The constraint on the $(m_t,\tan\beta)$ plane from the
SU(5) Yukawa unification condition (from Ref.~[15]).}\label{Figure2}
\vspace{7in}
\end{figure}

\begin{figure}[p]
\caption{Effect of ``threshold corrections" on the SU(5) Yukawa unification
condition for $m_b=4.25\GeV$ (from Ref.~[35]).}\label{Figure3}
\vspace{9.5in}
\end{figure}

\section{GUT Properties}
\subsection{SU(5)}
\begin{description}
\item[$\bullet$] GUT superpotential \cite{Dickreview}:
\beq
W_G=\lambda_1(\coeff{1}{3}\Sigma^3+\coeff{1}{2}M\Sigma^2)
+\lambda_2H(\Sigma+3M')\bar H
\eeq
where $\Sigma$ is the \r{24} of Higgs whose VEV
$\vev{\Sigma}=M\,{\rm diag}(2,2,2,-3,-3)$ breaks $SU(5)$ down to
$SU(3)\times SU(2)\times U(1)$, and $H=\{H_2,H_3\}$ is the Higgs pentaplet.
\item[$\bullet$] Doublet-triplet (2/3) splitting: The choice $M=M'$ makes the
triplet $H_3$ heavy, while keeping the doublet $H_2$ light. This fine-tuning is
avoided by the ``missing partner mechanism" where the \r{75} breaks the GUT
symmetry (instead of the \r{24}) and the couplings ${\bf50}\cdot{\bf75}\cdot
h$,
$\overline{\bf50}\cdot{\bf75}\cdot\bar h$ effect the 2/3 splitting. These
representations (\r{75},\r{50},\rb{50}) have seldom been considered in heavy
threshold analyses \cite{HY}. For other methods to solve this problem see
Ref.~\cite{Anselm}.
\item[$\bullet$] GUT fields: The heavy GUT fields, their transformation
properties under $SU(3)\times SU(2)$, their mass, and their usual notation are
given below
\bdm
\begin{array}{llll}
{\rm Field}&SU(3)\times SU(2)&{\rm Mass}&{\rm Name}\\
H_3,\bar H_3&(3,1),(\bar 3,1)&5\lambda_2M&M_H\\
\Sigma^8&(8,1)&\coeff{5}{2}\lambda_1M& M_\Sigma\\
\Sigma^3&(1,3)&\coeff{5}{2}\lambda_1M& M_\Sigma\\
\Sigma^0&(1,1)&\coeff{1}{2}\lambda_1M& ({\rm SM\ singlet})\\
X,Y&(3,2),(\bar 3,2)&5gM&M_V
\end{array}
\edm
One can see that only three mass parameters ($M_H,M_\Sigma,M_V$) are needed to
describe the heavy GUT fields, and thus the heavy threshold effects.
\item{$\bullet$} Proton Decay:\\

\noindent {\em Dimension-six}: is mediated by the $X,Y$ gauge bosons. The
largest mode is $p\to e^+\pi^0$, which if dominant would give
$\tau_p\sim3.3\times10^{35}(M_U/10^{16})^4\y$, thus it is basically
unobservable. However, the experimental bound on this mode implies
$M_U\gsim10^{15}\GeV$.\\

\noindent {\em Dimension-five}: is mediated by the Higgs triplet fields
$H_3,\bar H_3$ through the couplings $\lambda_u 10_f\cdot 10_f\cdot 5_h\supset
Q\cdot Q\cdot H_3$ and
$\lambda_d 10_f\cdot\bar 5_f\cdot\bar 5_h\supset Q\cdot L\cdot\bar H_3$
and the mixing term $\sim M_UH_3\bar H_3$. This operator needs to be ``dressed"
by a chargino loop, and the largest contribution comes from CKM mixing with the
second generation. The largest mode is $p\to\bar\nu_{\mu,\tau}K^+$ and a
schematic expression for this ``partial lifetime" is given by \cite{ANoldpd}
\beq
\tau(p\to\bar\nu_{\mu,\tau}K^+)\sim\left|M_H\sin2\beta{1\over f}
{1\over 1+y^{tK}}\right|^2.
\eeq
In this expression: (i) $M_H$ is the Higgs triplet mass: large $M_H$ makes the
lifetime longer; (ii) $\sin2\beta=2\tan\beta/(1+\tan\beta)$: small $\tan\beta$
needed to keep lifetime long enough; (iii) $f$: one-loop dressing function
which goes as $f\sim m_{\chi^\pm_1}/m^2_{\tilde q}$: heavy squarks and light
charginos are preferred; (iv) $1+y^{tK}$: ratio of third-to-second generation
contribution to dressing. Strong constraints on the parameter space of the
minimal $SU(5)$ supergravity model follow \cite{ANpd,LNP,LNPZ}. The
experimental constraint on the proton lifetime can be evaded by either: (a)
increasing the Higgs triplet mass, which is taken to be $M_H<(3-10)M_U$ in
these analyses (higher values produce too large heavy threshold corrections to
$\sin^2\theta_W$); (b) relaxing the naturalness constraint of $m_{\tilde
q,\tilde g}<1\TeV$. It must be emphasized that the values calculated this way
are {\em upper} bounds (since one uses an upper bound on $M_H$) and could well
be much larger if $M_H\approx M_U$ ($M_H\gsim M_U$ is required). In any event,
the next generation of proton decay experiments (SuperKamiokande and Icarus)
should carve out a large fraction of the remaining parameter space in this
model.
\end{description}

\subsection{SU(5)xU(1)}
\begin{description}
\item[$\bullet$] GUT superpotential \cite{EriceDec92}:
\beq
W_G=\lambda_4 HHh+\lambda_5\bar H\bar H\bar h+\lambda_6 F\bar H\phi
+\mu h\bar h
\eeq
where $H=\{Q_H,d^c_H,\nu^c_H\}$, $\bar H=\{Q_{\bar H},d^c_{\bar H},
\nu^c_{\bar H}\}$ are $SU(5)$ decaplets, $h=\{H_2,H_3\},\bar h$ are $SU(5)$
pentaplets, $\phi$ in an $SU(5)$ singlet, and the matter fields are in
$F=\{Q,d^c,\nu^c\}$, $\bar f=\{L,u^c\}$, and $l=e^c$. The vevs
$\vev{\nu^c_{H,\bar H}}$ break $SU(5)\times U(1)$
down to $SU(3)\times SU(2)\times U(1)$. This property, \ie, no need for adjoint
representations to break the GUT symmetry, is central to the appeal of
flipped $SU(5)$ as a string-derived model.
\item[$\bullet$] Doublet-triplet (2/3) splitting: no need for additional
representations
\bdm
\left.
\begin{array}{ll}
H\cdot H\cdot h&\supset\vev{\nu^c_H}d^c_H\,H_3\\
\bar H\cdot\bar H\cdot\bar h&\supset\vev{\nu^c_{\bar H}}d^c_{\bar H}\,\bar H_3
\end{array}\right\}
\begin{array}{l}
H_3,\bar H_3\ {\rm heavy}\\
H_2,\bar H_2\ {\rm light}
\end{array}
\edm

\item[$\bullet$] Proton decay:\\

\noindent{\em Dimension-six}: as in the minimal $SU(5)$ case.\\

\noindent{\em Dimension-five}: are suppressed relative to the $SU(5)$ case by
$\sim(M_Z/M_U)^2\lsim10^{-28}$, \ie, negligible. Reason: $H_3,\bar H_3$ mixing
is $\sim\mu\sim M_Z$, as opposed to $\sim M_U$ in $SU(5)$, while the Higgs
triplet mass is also $\sim M_U$.
\item[$\bullet$] Neutrino masses: the singlet field $\phi$ enlarges the usual
$2\times2$ see-saw matrix to $3\times3$. This mechanism is essential for the
consistency of the model, since otherwise the neutrinos would acquire large
masses
\bdm
\left.
\begin{array}{rll}
&F\cdot\bar H\cdot\phi&\to \vev{\nu^c_{\bar H}}\nu^c\,\phi\\
\lambda_u&F\cdot\bar f\cdot\bar h&\to m_u\nu\nu^c
\end{array}
\right\}
M_\nu=
\begin{array}{c}
\nu\\ \nu^c\\ \phi
\end{array}
\stackrel{\begin{array}{ccc} \nu\quad&\nu^c&\quad\phi\end{array}}
{\left(
\begin{array}{ccc}
0&m_u&0\\ m_u&0&M_U\\ 0&M_U&-
\end{array}
\right)}
\edm
The see-saw mechanism then gives $m_{\nu_{e,\mu,\tau}}\sim m^2_{u,c,t}/M^2_U$.
Incorporating all the appropriate renormalization group factors, it has
been shown that the $\nu_e,\nu_\mu$ sector could reproduce the needed MSW
effect in the solar neutrino flux \cite{chorus}, $\nu_\tau$ could be a
hot dark matter candidate \cite{ELNO}, and $\nu_\mu-\nu_\tau$ oscillations
could be observed at forthcoming experiments \cite{chorus}. Moreover, the
out-of-equilibrium decays of the ``flipped neutrinos" ($\nu^c$) could generate
a lepton asymmetry, which would later be processed into a baryon asymmetry
by electroweak non-perturbative interactions \cite{ENO}.
\end{description}

\section{Soft Supersymmetry Breaking}
\label{soft}
Spontaneous breaking of supergravity (\eg, induced dynamically by gaugino
condensation in the hidden sector) results in a global supersymmetric theory
{\em plus} a set of calculable soft supersymmetry breaking terms. Here
``soft" mean operators of dimension $\le3$ which do not regenerate the
quadratic divergences which supersymmetry avoids de facto. There are three
classes of such soft-supersymmetry-breaking terms:\\

\noindent$\bullet$ {\bf Gaugino masses} (parameters = 3)
\begin{description}
\item[--] $M_3,M_2,M_1$ for the three $SU(3)_C,SU(2)_L,U(1)_Y$ gauginos,
respectively.
\item[--] $SU(5)$ symmetry implies $M_3=M_2=M_1$.
\item[--] Universal soft-supersymmetry breaking decrees: $M_3=M_2=M_1=m_{1/2}$
at $M_U$
\item[--] This relation is almost universally adopted in low-energy
supersymmetric phenomenological studies (for exceptions see
Ref.~\cite{non-univ-gauginos}). For recent studies of two-loop effects on the
running of the $M_i$ see Ref.~\cite{Yamada}.
\end{description}
\noindent$\bullet$ {\bf Scalar masses} (parameters = $5\times3+2 = 17$)
\begin{description}
\item[--] $(\tilde Q,\tilde U^c,\tilde D^c,\tilde L,\tilde E^c)_i$, $i=1,2,3$
squarks and sleptons; $H_1,H_2$ Higgs bosons.
\item[--] Universality decrees: $m_{(\tilde Q,\tilde U^c,\tilde D^c,\tilde
L,\tilde E^c)_{1,2,3}}=m_{H_{1,2}}=m_0$ at $M_U$.
\item[--] Departures from universality are strongly constrained by
flavor-changing-neutral-current (FCNC) processes in the $K-\bar K$ system (most
notably the CP-violating $\epsilon$ parameter) \cite{EN}.
\item[--] Such departures are generic in string-inspired
supergravities \cite{IL,Casas}.
\end{description}

\noindent$\bullet$ {\bf Scalar couplings} (parameters = $3+1 = 4$)
\begin{description}
\item [--] To each superpotential coupling there corresponds one scalar
coupling:
\bdm
\begin{array}{ll}
\lambda_tQt^cH_2&\to \lambda_tA_t\tilde Q\tilde t^c H_2\\
\lambda_bQb^cH_1&\to \lambda_bA_b\tilde Q\tilde b^c H_1\\
\lambda_\tau L\tau^cH_1&\to \lambda_\tau A_\tau\tilde L\tilde\tau^c H_1\\
\mu H_1 H_2&\to \mu B H_1 H_2
\end{array}
\edm
\item[--] Universality decrees: $A_t=A_b=A_\tau=A$ at $M_U$.
\end{description}

\noindent$\bullet$ {\bf Some particular (string-inspired)
soft-supersymmetry-breaking scenaria}
\begin{description}
\item (a) No-Scale: $m_0=A=0$ \cite{LN};
\item (b) Strict No-scale: $m_0=A=0$ and $B(M_U)=0$;
\item (c) Dilaton: $m_0=\coeff{1}{\sqrt{3}}\,m_{1/2},\ A=-m_{1/2}$ \cite{KL};
\item (d) Special Dilaton: $m_0=\coeff{1}{\sqrt{3}}\,m_{1/2},\ A=-m_{1/2}$ and
$B(M_U)=\coeff{2}{\sqrt{3}}\,m_{1/2}$;
\item (e) Moduli: $m_0\sim m_{3/2}$, $m_{1/2}\sim(\alpha/4\pi)m_{3/2}$
(\ie, $m_{1/2}/m_0\ll1$) \cite{IL,Casas}.
\end{description}

\noindent All the soft-supersymmetry-breaking parameters, and the
gauge and Yukawa couplings evolve to low energies as prescribed by the
appropriate set of coupled RGEs. It is important to note that the values of
$\mu$ and $B$ do not feed into the other RGEs, and therefore they need not be
specified at high energies.

\noindent$\bullet$ {\bf Parameter count at low energies:}
\bdm
\begin{array}{lccc}
{\rm Parameter}&{\rm MSSM}&{\rm SUGRA}&\\
M_1,M_2,M_3&3&1&(m_{1/2})\\
(\tilde Q,\tilde U^c,\tilde D^c,\tilde L,\tilde E^c)_i&15&1&(m_0)\\
\tilde H_1,\tilde H_2&2&0&(m_0)\\
A_t,A_b,A_\tau&3&1&(A)\\
B&1&1&{\rm (determined\ by\ radiative}\\
\mu&1&1&{\rm electroweak\ breaking)}\\
\lambda_{b,t,\tau},\tan\beta&2&2&\\
&-&-&\\
{\rm Total:}&27&7&
\end{array}
\edm

\noindent The two minimization conditions of the electroweak scalar
potential (to be discussed in the next section) impose two additional
constraints which can be used to determine $\mu,B$ (at low energies)
and thus reduce the parameter count down to {\bf5} (versus 25 in the MSSM):
$m_t,\tan\beta,m_{1/2},m_0,A$. In the particular scenarios mentioned above, the
parameters are just three ($m_t,\tan\beta,m_{1/2}$), or even two
($m_t,m_{1/2}$) in scenarios (b) and (d).

\section{Radiative Electroweak Breaking}
We now discuss the mechanism by which the electroweak symmetry is broken.
In supergravity theories this occurs via radiative corrections in the presence
of soft-supersymmetry-breaking masses \cite{EWx}. This mechanism connects in a
nontrivial way various aspects of these theories, such as the physics at the
high and low scales, the breaking of supersymmetry, and the value of the
top-quark mass. In the MSSM electroweak symmetry breaking is put in by hand and
the top-quark mass plays no special role.

\subsection{Tree-level minimization}
The tree-level Higgs potential is given by
\begin{eqnarray}
V_0&=&(m^2_{H_1}+\mu^2)|H_1|^2+(m^2_{H_2}+\mu^2)|H_2|^2
				+B\mu(H_1H_2+{\rm h.c.})\nonumber\\
&&+\coeff{1}{8}g^2_2(H^\dagger_2{\bf\sigma}H_2+H^\dagger_1{\bf\sigma}H_1)^2
+\coeff{1}{8}g'^2\left(|H_2|^2-|H_1|^2\right)^2,
\end{eqnarray}
where $H_1\equiv{{H^0_1\choose H^-_1}}$ and $H_2\equiv{{H^+_2\choose H^0_2}}$
are the two complex Higgs doublet fields. Assuming that only the neutral
components get vevs, the expression for $V_0$ simplifies to
\beq
V_0=(m^2_{H_1}+\mu^2)h_1^2+(m^2_{H_2}+\mu^2)h_2^2+2B\mu h_1h_2
+\coeff{1}{8}(g_2^2+g'^2)(h^2_2-h^2_1)^2,
\eeq
where $h_i= {\rm Re}\,H^0_{1,2}$. One can then write down the minimization
conditions $\partial V_0/\partial h_i=0$ and obtain
\begin{eqnarray}
\mu^2&=&{m^2_{H_1}-m^2_{H_2}\tan^2\beta\over\tan^2\beta-1}
-\coeff{1}{2}M^2_Z,\label{mu2}\\
B\mu&=&-\coeff{1}{2}\sin2\beta(m^2_{H_1}+m^2_{H_2}+2\mu^2).
\end{eqnarray}
The solutions to these equations will be physically sensible only if they
reflect a minimum away from the origin
\beq
{\cal S}=(m^2_{H_1}+\mu^2)(m^2_{H_2}+\mu^2)-B^2\mu^2<0\label{stab}
\eeq
of a potential bounded from below
\beq
{\cal B}=m^2_{H_1}+m^2_{H_2}+2\mu^2+2B\mu>0.
\eeq
Taking the second derivatives $\partial^2 V_0/\partial h_i\partial h_j$ one
can determine the tree-level masses of the five physical Higgs bosons: the
CP-even states $h,H$, the CP-odd state $A$, and the charged Higgs boson
$H^\pm$. An important result is that the lightest Higgs boson mass is bounded
above $m_h\le|\cos2\beta|M_Z$, although one-loop corrections relax this
constraint considerably. Also, $m_h<m_A$, $m_H>M_Z$, and $m_{H^\pm}>M_W$, which
may also be affected by radiative corrections.

\subsection{One-loop minimization}
The minimization of the RGE-improved tree-level Higgs potential described
above suffers from a scale-dependence problem. That is, the physical output
obtained depends considerably on the scale one chooses to perform the
minimization, \ie, the scale at which the RGEs are stopped. This problem
is most simply stated as
\beq
{dV_0\over d\ln Q}\not=0,
\eeq
that is, the tree-level Higgs potential does not satisfy the renormalization
group equation. In practice, as the scale is lowered one typically has the
sequence of events pictured below
\vspace{1cm}
\begin{figure}[h]
\vbox{\LARGE
\noindent$Q>Q_0$\qquad ${\cal S}>0$ \qquad\qquad\qquad\qquad\qquad vevs=0\\

\vspace{1.5cm}
\noindent$Q=Q_0$\qquad ${\cal S}=0$ \qquad\qquad\qquad\qquad\qquad
vevs$\approx0$\\

\vspace{1.5cm}
\noindent$Q<Q_0$\qquad ${\cal S}<0$ \qquad\qquad\qquad\qquad\qquad vevs ok\\

\vspace{1.5cm}
\noindent$Q<Q_1$\qquad ${\cal B}<0$ \qquad\qquad\qquad\qquad\qquad
vevs$\to\infty$\\
}
\end{figure}

\noindent That is, the tree-level potential has a minimum at the origin for
scales $Q>Q_0$, then it develops at minimum away from the origin (${\cal S}<0$)
with vevs which grow to be such that
$M^2_Z=\coeff{1}{2}(g^2+g'^2)(v^2_1+v^2_2)$. However, for scales $Q<Q_1$ this
minimum becomes unbounded from below (${\cal B}<0$) and the vevs run away to
infinity. Thus, the vevs vary a lot for scales $Q\lsim1\TeV$ \cite{GRZ}, as
shown schematically in the following figure:
\begin{figure}[t]
\vspace{2.5in}
\end{figure}

\vspace{2cm}
 The solution to this problem is to use the one-loop effective potential
$V_1=V_0+\Delta V$, which satisfies ${dV_1\over d\,\ln Q}=0$ (to one-loop
order), where
\beq
\Delta V=\coeff{1}{64\pi^2}{\rm Str}\,{\cal M}^4\left(\ln{{\cal M}^2\over Q^2}
-\coeff{3}{2}\right)
\eeq
with ${\rm Str}{\cal M}^2=\sum_j(-1)^{2j}(2j+1){\rm Tr}{\cal M}^2_j$.
Following this procedure the vevs are $Q$-independent (up to two-loop effects)
in the range of interest ($\lsim1\TeV$) \cite{GRZ}. However, one must perform
the minimization numerically (a non-trivial task) and {\em all} the spectrum
enters into $\Delta V$ (although $\tilde t,\tilde b$ are the dominant
contributions) (see \eg, \cite{aspects}). This procedure also gives
automatically the {\em one-loop corrected} Higgs boson masses (taking second
derivatives of $V_1$).

\subsection{Radiative Symmetry Breaking}
To grasp the concept most easily, let us consider the simple (although
unrealistic) case of $\mu=0$. Let us also not worry about the one-loop
correction to the Higgs potential. Neither of these simplifications will
affect the physical mechanism which we want to illustrate. In this case the
``stability" condition becomes
\beq
{\cal S}\to m^2_{H_1}\cdot m^2_{H_2}<0.
\eeq
This means that one must arrange that one $m^2_{H_i}<0$ somehow. No help is
available from low-energy physics inputs alone (\ie, this is put in by hand in
the MSSM). To proceed, consider RGEs for the (first- and second-generation)
scalar masses schematically (setting $\lambda_b=\lambda_\tau=0$)
\beq
{d\widetilde m^2\over dt}={1\over(4\pi)^2}\left\{
-\sum_i c_i g^2_i M^2_i + c_t\lambda^2_t\left(\sum_i\widetilde
m^2_i\right)\right\},
\eeq
where the various coefficients are given below
\bdm
\begin{array}{cccc}
&c_t&c_3&c_2\\
H_1&0&0&6\\
H_2&6&0&6\\
\widetilde Q&0&\coeff{32}{3}&6\\
\widetilde U^c&0&\coeff{32}{3}&0\\
\widetilde D^c&0&\coeff{32}{3}&0\\
\widetilde L&0&0&6\\
\widetilde E^c&0&0&0
\end{array}
\edm
The result of running these RGEs is illustrated below for the indicated values
of the parameters.
\begin{figure}[h]
\vspace{3.5in}
voffset=-120}
\end{figure}
\vspace{2cm}
Note that $m^2_{H_2}<0$ while $m^2_{H_1}>0$ for $Q<Q_0$. Since one can show
that $\mu$ is small in this case ($\mu\approx30\GeV$), this implies that ${\cal
S}<0$, \ie, that a minimum away from the origin has developed. If $\mu$ were
not small, radiative breaking will require a large enough negative value for
$m^2_{H_2}$ (in order to possibly get ${\cal S}<0$, see
Eqs.~(\ref{mu2},\ref{stab})). This will generally be the case in models where
$m_0\ll m_{1/2}$. The top-quark Yukawa coupling ($\lambda_t$) plays a
{\em fundamental} role in driving $m^2_{H_2}$ to negative values. However, this
is only possible if it is large enough to counteract the effect of the gauge
couplings. This is why early on people said that this mechanism required a
``heavy top quark". Nowadays, any allowed top-quark mass is heavy enough.
Note also that $m^2_{\tilde Q,\tilde U^c,\tilde D^c}>0$ because of the large
$\alpha_3$ contribution to their running. For the same reason the sleptons
($\tilde L,\tilde E^c$) renormalize much less.

\subsection{Constraints on the $(m_t,\tan\beta)$ plane}
As we have seen above, in supersymmetric unified models with radiative
electroweak symmetry breaking, there are only five parameters: three
soft-supersymmetry-breaking parameters ($m_{1/2},m_0,A$) and $m_t,\tan\beta$.
It turns out that the two-dimensional area spanned by the last two parameters
is completely bounded \cite{noscale,aspects} using the tree-level or one-loop
minimization procedures discussed above. This is not the case for any
other pair of variables. The shape of the resulting boundaries depends on the
values of the soft-supersymmetry-breaking parameters, the sign of $\mu$,
and the phenomenomenological constraints which are imposed on the spectrum.

First let us study the case of {\em no phenomenological constraints}. In the
figures below, solid (dashed) lines denote the allowed boundaries obtained by
enforcing radiative breaking using the tree-level (one-loop) Higgs potential.
We have taken $m_0=A=0$ (the no-scale scenario) and $m_{1/2}=150,250\GeV$.

\begin{figure}[h]
\vspace{4.7in}
\end{figure}
\clearpage

The bounded region is seen to have five boundaries:
\begin{description}
\item (i) Top boundary: this is the slanted line which limits the magnitude of
$\tan\beta$. Along this line the CP-odd Higgs boson mass vanishes ($m_A=0$) and
${\cal B}=0$, \ie, the potential becomes unbounded for points above this line.
Also, on this boundary the relation $\lambda_b\approx\lambda_t$ holds, which
implies that $\tan\beta\approx m_t/m_b(M_Z)\approx m_t/3.77$.
\item (ii) Upper corner: the largest allowed value of $\tan\beta$ is determined
by the fixed point in $\lambda_b$ (as discussed in Sec.~\ref{fixed}); this
value is $m_b(m_b)$ dependent. The rounded portion at the top and towards
the larger values of $m_t$ results from the strengthening of the fixed point
bound because of the non-zero top-quark Yukawa coupling.
\item (iii) Right boundary: this is determined by the fixed point in
$\lambda_t$ and entails a $\tan\beta$-dependent upper bound on $m_t$, which
is quite restrictive for low $\tan\beta$ (see Sec.~\ref{fixed}).
\item (iv) Bottom boundary: here $\tan\beta>1$ which is a direct and important
consequence of the radiative breaking mechanism. This lower bound is routinely
assumed in phenomenological analyses and has no other known explanation.
\item (v) Left boundary: on this line $\mu$ vanishes, \ie, to the left of
the line $\mu^2<0$. Moreover, the one-loop procedure yields the largest
correction to the tree-level result precisely on this boundary. For example,
for $\mu>0$ and $m_{1/2}=250\GeV$, on the one-loop left boundary
$\mu_{loop}=0$, whereas $\mu_{tree}\approx60\GeV$.
\end{description}
\clearpage

For comparison, we now consider the case of $m_0=m_{1/2}$ and $A=0$, to
study the effect of the soft-supersymmetry-breaking parameters on the shape
of the boundary. (Still no phenomenological constraints have been applied.)
The thing to note are the larger $\mu$ values which are generated by the larger
values of $m^2_{H_{1,2}}$ in the expression for $\mu^2$. The position of the
left boundary has also shifted (to the right).

\begin{figure}[h]
\vspace{5in}
\end{figure}
\clearpage

Now let us include the various {\em phenomenological cuts}: (i) the
lightest supersymmetric particle (LSP) should be neutral, in fact, one demands
that it be the lightest neutralino; (ii) the chargino mass is bounded below
by LEP, $m_{\chi^\pm_1}>45\GeV$; (iii) and so are the slepton masses,
$m_{\tilde l}>45\GeV$; (iv) the neutralino contributions to $\Gamma_Z$ should
be small enough; (v) the lightest Higgs boson mass is also bounded below by
LEP, $m_h>43\GeV$.

The resulting allowed region in the $(m_t,\tan\beta)$ plane for the case
$m_0=A=0$ is shown below.
The most notable change is the more restricted range of $\tan\beta$
values which are allowed: the maximum allowed $\tan\beta$ is decreased and
for $m_t\gsim 100\GeV$ this value decreases with $m_t$. It is interesting to
note how large a constraint it is to have a lower bound on the top-quark mass
which is ever increasing ($m_t\gsim110\GeV$ at present). In this figure one can
also see the effect of $A$ (dotted line: $A=m_{1/2}$; dashed line:
$A=-m_{1/2}$).

\begin{figure}[h]
\vspace{5.5in}
\end{figure}
\clearpage

To conclude we show below the case of $m_0=m_{1/2}$ when the
phenomenological constraints are imposed. Note how much weaker these  cuts
become. This is simply because the spectrum is heavier than in the previous
(no-scale) case.

\begin{figure}[h]
\vspace{5.25in}
\end{figure}

Without further assumptions about the soft-supersymmetry-breaking parameters
(such as those mentioned in Sec.~\ref{soft}) or more phenomenological
constraints (\eg, proton decay), the five-dimensional parameter space is still
rather vast. However, this situation changes drastically in specific models,
such as the no-scale \cite{LNZI} and dilaton \cite{LNZII} flipped $SU(5)$
supergravity models, where $m_0$ and $A$ are functions of $m_{1/2}\propto
m_{\tilde g}$. Furthermore, more restricted versions of each of these models
allow to determine $\tan\beta$ as a function of $m_t$ and $m_{1/2}$ and a
two-dimensional parameter space results. In this case, the allowed areas in
$(m_t,\tan\beta$) space shown above degenerate into a line on those figures.
Such two- or three-parameter models have been shown to be quite predictive,
and the expectations for collider processes at the Tevatron \cite{LNWZ},
LEP \cite{LNPWZh,LNPWZ}, and HERA \cite{hera} have been explicitly calculated,
as well as indirect probes through one-loop precision electroweak corrections
at LEP \cite{ewcorr}, the FCNC $b\to s\gamma$ rare decay at CLEO
\cite{bsgamma+bsg-eps}, the anomalous magnetic moment of the muon \cite{g-2},
and also the prospects for cosmological dark  matter \cite{KLNPYdm,LNZI,LNZII},
and indirect neutralino detection at neutrino telescopes \cite{NT}. Some of
these calculations have also been performed in the minimal $SU(5)$ supergravity
model, where the constraints from proton decay are so powerful that the general
three-dimensional soft-supersymmetry-breaking parameter space becomes quite
manageable \cite{ANpd,LNP,LNPZ,ANcosm,poles}. For recent reviews of this line
of research see \eg, Ref.~\cite{EriceJul92,Dick'sTalk}.

\section{Conclusions}

In this lecture I have shown that supersymmetric unified theories with
radiative electroweak symmetry breaking are highly predictive models of
low-energy supersymmetry, once all necessary elements are incorporated. As such
they stand in sharp contrast with the minimal supersymmetric standard model,
whose ``minimality" in the field content is all but washed out by the large
number of unknown parameters needed to describe it. I would also like to
emphasize that one must consider all aspects of any such unified models before
being able to judge their experimental viability. It is not unusual for various
sectors of the theory to complement each other as far as constraints on the
parameter space are concerned. Finally, building on the structure I have
described, it is possible to construct well motivated (and more constrained)
models which can be used to calculate all sparticle and Higgs boson masses, as
well as all conceivable observables of interest at present and future
experimental facilities. These computations and their intricate correlations
should play a very important role in accumulating direct and indirect evidence
for the supersymmetric model which is actually realized in nature.

\section*{Acknowledgements}
I would like to thank the Ettore Majorana Center for Scientific Culture and
Professor Zichichi for their kind invitation to deliver this Lecture. This
work has been supported by an SSC Fellowship.

\end{document}